\documentclass{PoS}

\title{Theoretical Prospects for B Physics}

\ShortTitle{Theoretical Prospects for B Physics}

\author{\speaker{Robert Fleischer}\\
        Nikhef, Science Park 105, 1098 XG Amsterdam, Netherlands\\
        \mbox{Department of Physics and Astronomy, VU University Amsterdam, 
        1081 HV Amsterdam, Netherlands}\\
        E-mail: \email{Robert.Fleischer@nikhef.nl}}

\abstract{The exploration of $B$-meson decays has reached an unprecedented level of 
sophistication, with a phase of even much higher precision ahead of us thanks to run 2
of the LHC and the future era of Belle II and the LHCb upgrade. For many processes, 
the theoretical challenge in the quest to reveal possible footprints of physics beyond 
the Standard Model will be the control of uncertainties from strong interactions.
After a brief discussion of the global picture emerging from the LHC data, I will focus on
the theoretical prospects and challenges for benchmark $B$ decays to search for 
new sources of CP violation, and highlight future opportunities to probe the 
Standard Model with strongly suppressed rare $B$ decays.}

\FullConference{Flavor Physics \& CP Violation 2015\\
		May 25-29, 2015\\
		Nagoya, Japan}

\begin{document}
\section{Introduction}
The data collected at the Large Hadron Collider (LHC) at CERN have led to many exciting 
results \cite{ellis}. In the search for new particles, the Higgs boson was eventually 
detected at the ATLAS and CMS detectors. It will be exciting to explore the properties of the 
Higgs in more detail in the future. In contrast to the discovery of the Higgs boson, no new 
particles originating from physics beyond the Standard Model (SM) were found 
at the high-energy frontier so far. Concerning the high-precision frontier, which is 
the domain of flavour -- in particular $B$ -- physics, there was impressive progress in 
studies of CP violation and rare decays at the LHC, where the key role was played by the 
dedicated LHCb experiment, complemented by studies at ATLAS and CMS. 

The LHC results have triggered a lot of interest in the theory community, which 
is reflected by many analyses in the recent literature. The measured flavour observables were 
found to be globally consistent with the SM picture although ``tensions" with respect to the SM 
have recently emerged in a variety of rare $B$-decay observables. These exciting phenomena, 
which are unfortunately not yet conclusive, offered hot topics for presentations and discussions 
at this conference. 

The LHC data have the following implications for the general structure of new physics (NP): 
we may encounter a large characteristic NP scale $\Lambda_{\rm NP}$, or (and?) symmetries 
may prevent large NP effects in the flavour sector and processes originating from flavour-changing 
neutral currents (FCNCs). The most prominent example of scenarios with the latter feature is 
given by NP models with ``Minimal Flavour Violation" (MFV), where the flavour and CP violation 
is -- sloppily speaking -- the same as in the SM. For a more detailed discussion, see
\cite{walt}. 

New perspectives both for direct NP searches at  ATLAS and CMS and for indirect searches in 
high-precision flavour analyses will arise at the recently started run 2 of the LHC with almost 
the double centre-of-mass energy of 13 TeV of the colliding protons. The LHC energy is obviously 
the key parameter for the production of new particles at this collider. However, also the production of 
$B$ mesons, which are the key players in flavour physics research, is almost doubled thanks to the 
higher energy of the LHC. There is a fruitful interplay between LHC and flavour physics, as discussed
in more detail in \cite{JorgeMC}.

\section{Theoretical Framework}
In theoretical analyses of $B$ decays, a huge hierarchy of scales arises, ranging from NP scales
$\Lambda_{\rm NP}\sim 10^{(0 ... ?)} \,\mbox{TeV}$ over the electroweak scale 
$\Lambda_{\rm EW}\sim 10^{-1} \,\mbox{TeV} $ to long-distance scales 
$\Lambda_{\rm QCD}\sim 10^{-4} \,\mbox{TeV}$, which are 
related to hadronic bound state effects of strong interactions described by QCD. 
Effective field theories offer the appropriate framework to deal with this situation:
the heavy degrees of freedom (NP particles,  top quark,  $Z$ and $W$ bosons) are ``integrated out" 
from appearing explicitly and are encoded in short-distance loop functions, perturbative QCD corrections
can be calculated in a systematic way, with renormalisation group techniques allowing the summation
of large logarithms $\log(\mu_{\rm SD}/\mu_{\rm LD})$. 
This machinery was successfully applied to the SM
and various NP scenarios \cite{BG}. The transition amplitude of a decay $\bar B\to\bar f$
is given as the matrix element of a low-energy effective Hamiltonian ${\cal H}_{\mbox{{\scriptsize eff}}}$.
In the SM, it takes the following general form \cite{BLO}:
\begin{equation}\label{SM-ampl}
\langle \bar f|{\cal H}_{\mbox{{\scriptsize eff}}}|\bar B\rangle=
\frac{G_{\rm F}}{\sqrt{2}}\sum_{j}\lambda_{\rm CKM}^{j}\sum_{k}
C_{k}(\mu)\,\langle \bar f |Q^{j}_{k}(\mu)|\bar B\rangle,
\end{equation}
where $G_{\rm F}$ is Fermi's constant and the $\lambda_{\rm CKM}^{j}$ are products
of CKM matrix elements; the short-distance physics is described by the Wilson coefficient $C_k(\mu)$, 
which can be calculated in perturbation theory. In the presence of NP, the $C_k(\mu)$ may 
get new contributions -- also with CP-violating phases -- and new operators may arise. 

The long-distance contributions to the transition amplitude in (\ref{SM-ampl}) 
are encoded in the hadronic matrix elements $\langle \bar f |Q^{j}_{k}(\mu)|\bar B\rangle$. 
These non-perturbative quantities 
limit the theoretical precision of flavour physics observables. However, there has recently been 
impressive progress in lattice QCD for the calculation of $B$ decay constants and form factors which 
enter the rare $B^0_{s,d}\to\mu^+\mu^-$ and semileptonic decays \cite{Khadra}. For a detailed 
overview, the reader is referred to the Flavour Lattice Averaging Group (FLAG) \cite{FLAG}, and
the discussion in \cite{AV}. 

Concerning the theoretical description of non-leptonic $B$ decays, QCD factorisation (QCDF) \cite{QCDF}, 
the perturbative QCD (PQCD) approach \cite{PQCD}, Soft Collinear Effective Theory (SCET) \cite{SCET},
and applications of QCD sum rules \cite{QCDSR} offer interesting frameworks. There has been technical 
progress, such as the recent calculation of the two-loop current--current operator contribution to the 
QCD penguin amplitude \cite{BBHL}. Nevertheless the theoretical description of non-leptonic $B$
decays remains generally a challenge, also in view of patterns in the experimental data. 
For a further theoretical discussion of non-leptonic $B$ decays, see \cite{sinha}.

Non-leptonic $B$ decays play the key role for the exploration of CP violation as the corresponding
CP asymmetries are generated through interference effects which show up in such transitions. 
Fortunately, the hadronic matrix elements cancel in certain observables or can be determined 
with the help of amplitude relations between various processes \cite{RF-rev}: there are methods 
using exact relations, others neglecting certain contributions -- typically from penguin topologies -- and 
strategies exploiting the flavour symmetries of strong interactions. In the future era of measurements 
with even higher precision, it will be essential to get a handle on the corresponding theoretical 
uncertainties, with the goal to match the theoretical with the experimental precision.

\section{Studies of CP Violation}
In the SM, flavour and CP violation is described by the Cabibbo--Kobayashi--Maskawa (CKM) matrix
\cite{Kobayashi}, which can be illustrated by the Unitarity Triangle (UT) with its angles $\phi_1\equiv\beta$, 
$ \phi_2\equiv\alpha$ and $\phi_3\equiv\gamma$. A variety of flavour-physics observables can be converted 
into constraints for the apex of the UT, utilising a fruitful interplay between theory and experiment. 
Over the last 15 years, since the start of the $e^+e^-$ $B$ factories with the BaBar and Belle detectors
at SLAC and KEK, respectively, we have seen impressive progress in the determination of the UT, 
which is reflected by the continuously updated analyses by the CKMfitter \cite{CKMfitter} and 
UTfit \cite{UTfit} collaborations. 

\subsection{Determination of $\gamma$}
Looking at the UT angles, $\gamma$ has the largest uncertainty. Using decays 
of the kind $B\to D^{(*)}K^{(*)}$ and $B_s\to D_s^\mp K^\pm$, which receive only 
contributions from tree-diagram-like topologies, $\gamma$ can be determined in a theoretically 
clean way \cite{CKM-10,MR}, where in the latter case of the $B^0_s$ modes $B^0_s$--$\bar B^0_s$
mixing is exploited. In the corresponding strategies, simply speaking, the hadronic matrix elements 
cancel in the corresponding observables. Moreover, these modes are very robust with respect to 
NP contributions and therefore offer reference determinations of the SM value of $\gamma$. 
The current data for the $B\to D^{(*)}K^{(*)}$ modes yield 
\begin{equation}\label{gamma-det}
\gamma=(73.2^{+6.3}_{-7.0})^\circ~\mbox{[CKMfitter]}, \quad 
\gamma=(68.3 \pm 7.5)^\circ~\mbox{[UTfit]}.
\end{equation}
In the era of Belle II \cite{Abe:2010gxa} and the LHCb upgrade \cite{LHCb-implications}, 
an experimental uncertainty of $\Delta\gamma_{\rm exp}\sim 1^\circ$ is expected, which is a
very impressive and exciting perspective. 

Decays with loop contributions offer also strategies to determine $\gamma$. Here $B_{(s)}\to \pi\pi,\pi K,KK$
decays play the key role, where amplitude relations following from the $SU(3)$ flavour symmetry of 
strong interactions can be complemented with QCDF/SCET/PQCD calculations of 
$SU(3)$-breaking corrections. The goal is to compare the value of $\gamma$ following from the pure 
tree decays with the value of $\gamma$ extracted by means of decays with loop contributions. The
central question related to these analyses is whether discrepancies between the tree and loop 
determinations of $\gamma$ will eventually show up, which would indicate new sources of CP violation. 
 
The most promising system for the latter kind of strategies involving loop topologies is given 
by the $B^0_s\to K^+K^-$, $B^0_d\to \pi^+\pi^-$ decays, where the hadronic parameters can be 
related to one another with the help of the $U$-spin symmetry of strong interactions 
\cite{RF-BsKK-99,RF-BsKK-07}. Using data for branching ratios and CP violation, 
$\gamma=(67.7^{+4.5}_{-5.0}|_{\rm input}\mbox{}^{+5.0}_{-3.7}|_{\rm {\it U} \, spin})^\circ$ was
extracted, where the uncertainties due to the input quantities and $U$-spin-breaking corrections were
made explicit \cite{RF-BsKK-07,FK-BsKK,Rob-PhD}. 
This result is in impressive agreement with the pure $B\to D^{(*)}K^{(*)}$ 
tree decay results in (\ref{gamma-det}). 

An interesting variant of this strategy was proposed in \cite{CFMS}. It combines the 
$B^0_s\to K^+K^-$, $B^0_d\to\pi^+\pi^-$ $U$-spin method \cite{RF-BsKK-99} with the Gronau--London 
isospin analysis of the $B\to \pi\pi$ system \cite{GL}. The LHCb collaboration extracted 
$\gamma=( 63.5^{+7.2}_{-6.7})^\circ$ (in good agreement with the result given above) and 
$\phi_s\equiv-2\beta_s = -(6.9^{+9.2}_{-8.0})^\circ$ for the $B^0_s$--$\bar B^0_s$ mixing phase 
\cite{LHCb-BsKK}, including also the first measurements of CP violation in the
$B^0_s\to K^+K^-$ channel, which have still large uncertainties. The combined analysis of $\gamma$ 
is more robust with respect to $U$-spin-breaking effects. However, for values of an $U$-spin-breaking 
parameter $\kappa$ smaller than 0.5, the differences with respect to the original 
$B_s\to K^+K^-$, $B_d\to\pi^+\pi^-$ strategy are very small. An interesting feature is 
the determination of $\phi_s$ \cite{RF-BsKK-07,FK-BsKK}, which turns out to be particularly stable 
with respect to $U$-spin-breaking effects. 

Yet another variant of the $U$-spin method has recently been proposed in \cite{BPPP}, 
applying it to $B\to PPP$ decays. It employs $B^0_{d,s}\to K_{\rm S}h^+h^-$ $(h=K,\pi)$ decays,
where time-dependent Dalitz plot analyses allow the measurement of the corresponding branching 
ratios and CP asymmetries. The $U$-spin method to extract $\gamma$ can be applied -- in analogy 
to the $B^0_s\to K^+K^-$, $B^0_d\to\pi^+\pi^-$ system -- at each point of the Dalitz plot. A potential 
advantage of using three-body decays is that the effects of $U$-spin breaking may be reduced 
by averaging over the Dalitz plot.

\subsection{CP Violation in $b\to s$ Penguin-Dominated Modes}
There is plenty of data on $B$ decays into CP eigenstates that are dominated by $b\to s$ penguin processes
\cite{HFAG}. The direct and mixing-induced CP asymmetries of these modes may encode contributions
from physics beyond the SM, but due to currently large uncertainties such effects cannot be resolved.
The future challenge is the control of the hadronic uncertainties. 

A particularly interesting decay in this respect is given by $B^0\to \pi^0 K^0$ \cite{FJPZ,GR}. In this case, 
there is an isospin relation between the amplitudes of the neutral $B\to \pi K$ modes, which implies 
a correlation between the CP asymmetries of the $B^0\to \pi^0 K_{\rm S}$ channel:
\begin{equation}\label{iso-rel}
\sqrt{2}\,A(B^0\to\pi^0K^0)\,+\,A(B^0\to\pi^-K^+)
=-\underbrace{\left[(\hat T+\hat C)e^{i\gamma}\,+
\,\hat P_{\rm ew}\right]}_{\mbox{$(\hat T+ \hat C)(e^{i\gamma}-q e^{i\omega})$}}
\equiv 3  A_{3/2}.
\end{equation}
Here the $A_{3/2}$ amplitude can be fixed through the $B^+\to\pi^+\pi^0$ branching ratio with the help 
of the $SU(3)$ flavour symmetry. This quantity is well behaved within QCD factorization, which allows
the inclusion of $SU(3)$-breaking corrections. The data for the CP-violating $B^0\to \pi^0 K_{\rm S}$ 
asymmetries show an intriguing picture with respect to the SM correlation \cite{FJPZ}. The current 
uncertainties are unfortunately too large to draw definite conclusions. The decay $B^0\to \pi^0 K^0$ 
and the analysis proposed in \cite{FJPZ} is an interesting playground for the Belle II experiment. 

As electroweak penguins, which are encoded in the parameter $q e^{i\omega}$ in (\ref{iso-rel}), 
have a significant impact, NP in this sector may resolve this situation. Such kind of physics beyond 
the SM is currently a hot topic in view of the
$B^0_d\to K^{*0}\mu^+\mu^-$ ``anomaly" discussed below, 
which could be accommodated, for instance, through 
models with extra $Z'$ bosons. In view of this situation, also other non-leptonic $B$-meson decays 
with sensitivity to electroweak penguins are put into the spotlight:
\begin{displaymath}
B^+\to \pi^0K^+,\, B^0_s\to \phi\phi,\, B^0_s\to \pi^0\phi,\, B^0_s\to \rho^0\phi, 
\end{displaymath}
These modes and possible signs of NP in the corresponding $B$-factory data were hot 
topics quite some time ago (see, e.g., \cite{EWP-rev,BFRS,BEJLLW}). It will be interesting 
to monitor these modes in the future.

\subsection{Precision Measurements of the $B^0_q$--$\bar B^0_q$  Mixing Phases}
Interference effects between $B^0_q$--$\bar B^0_q$ mixing and decay processes give rise
to mixing-induced CP violation \cite{RF-rev}. These phenomena involve the CP-violating phases
\begin{equation}
\phi_d=2\beta+\phi_d^{\rm NP}, \quad \phi_s=-2\delta\gamma+\phi_s^{\rm NP},
\end{equation}
where the SM contributions originating from box topologies are given by the UT angle $\beta$ 
and $\delta\gamma=\lambda^2\eta$; $\lambda$ and $\eta$ are Wolfenstein 
parameters \cite{wolf}. The decay $B^0_d\to J/\psi K_{\rm S}$ plays the central role for the determination 
of  $\phi_d$ while the $B^0_s\to J/\psi \phi$ channel gives access to the phase $\phi_s$. In the 
presence of CP-violating NP contributions to $B^0_q$--$\bar B^0_q$ mixing, which may also arise at
the tree level, the phases $\phi_q^{\rm NP}$ differ from zero \cite{BaFl}.

If penguin contributions to $B^0_d\to J/\psi K_{\rm S}$ and $B^0_s\to J/\psi \phi$  decays are neglected,
the hadronic matrix elements cancel in the mixing-induced CP asymmetries, thereby allowing the 
extraction of $\phi_d$ and $\phi_s$ (for a detailed discussion, see \cite{RF-rev}). In view of the agreement
of the current measurements with the SM and the future era of experimental high-precision analyses, 
the penguin contributions, which are doubly Cabibbo-suppressed, have to be controlled 
\cite{RF-psiKS}
--\cite{FNW}.

\subsubsection{The $B^0_d\to J/\psi K_{\rm S}$, $B^0_s\to J/\psi K_{\rm S}$ System}
The decay amplitude of the $B^0_d\to J/\psi K_{\rm S}$ channel takes the following form:
\begin{equation}
A(B_d^0\to J/\psi\, K_{\rm S})=\left(1-\lambda^2/2\right){\cal A'}
\left[1+\epsilon a'e^{i\theta'}e^{i\gamma}\right],
\end{equation}
where the parameter $a'e^{i\theta'}$ with $\epsilon\equiv\lambda^2/(1-\lambda^2)\sim0.05$ describes 
contributions from doubly Cabibbo-suppressed penguin topologies \cite{RF-psiKS}, which are usually
neglected. The general expression for the mixing-induced CP asymmetry taking them into account
is given by
 \begin{equation}
\frac{S(B_d\to J/\psi K_{\rm S})}{\sqrt{1-C(B_d\to J/\psi K_{\rm S})^2}}
=\sin(\phi_d+\Delta\phi_d),
\end{equation}
where the hadronic phase shift $\Delta\phi_d$ is fixed through 
\begin{equation}
\sin\Delta\phi_d\propto 2 \epsilon a'\cos\theta' \sin\gamma+\epsilon^2a'^2
\sin2\gamma
\end{equation}
\begin{equation}
\cos\Delta\phi_d\propto1+ 2 \epsilon a'\cos\theta' \cos\gamma+\epsilon^2a'^2
\cos2\gamma,
\end{equation}
characterising the impact of the penguin contributions \cite{FFJM}. 

At Belle II and the LHCb upgrade, the experimental precision requires the control of the 
penguin corrections to reveal possible CP-violating NP contributions to $B^0_d$--$\bar B^0_d$ mixing. 
The hadronic phase shift $\Delta\phi_d$ cannot be calculated in a reliable way. On the other hand,
it can be determined with the help of experimental data for ``control channels". 

The most prominent example is given by $B^0_s\to J/\psi K_{\rm S}$ \cite{RF-psiKS}. Its key 
feature is the following structure of the decay amplitude:
\begin{equation}
A(B^0_s\to J/\psi K_{\rm S})\propto\left[1-a e^{i\theta}e^{i\gamma}\right],
\end{equation}
where the parameter $a e^{i\theta}$, which is the counterpart of $a'e^{i\theta'}$, is not suppressed
by the tiny $ \epsilon$. The $U$-spin symmetry implies the relation
\begin{equation}
a e^{i\theta} = a'e^{i\theta'},
\end{equation}
thereby allowing the control of the penguin effects in the determination of $\beta$ from the CP violation
in the $B^0_d\to J/\psi K_{\rm S}$ channel \cite{RF-psiKS,KR-pen}. 

Using the $SU(3)$ flavour symmetry with further dynamical assumptions on the neglect of 
exchange and penguin annihilation topologies, a global analysis of the data on $B\to J/\psi X$
decays yields the following results \cite{KR-pen}:
\begin{equation}\label{a-det}
a = 0.19^{+0.15}_{-0.12} \:,\qquad \theta = \left(179.5 \pm 4.0\right)^{\circ}\:,\qquad 
\phi_d = \left(43.2^{+1.8}_{-1.7}\right)^{\circ}.
\end{equation}
The corresponding hadronic phase shift 
$\Delta\phi_d^{\psi K_{\rm S}^0} = -\left(1.10^{+0.70}_{-0.85}\right)^{\circ}$ gives guidance 
for the importance of penguin effects. Internal consistency checks of the $B\to J/\psi X$ data 
look fine and do not indicate any anomalous features. 

In the future, there will be exciting prospects to go beyond this global fit analysis by means of precision 
measurements of CP violation in $B^0_s\to J/\psi K_{\rm S}$. The results in (\ref{a-det}) correspond
to the following SM predictions: 
\begin{equation}
\begin{array}{rcl}
{\cal A}_{\rm CP}^{\rm dir}(B_s\to J/\psi K_{\mathrm S}^0) |_{\rm SM}
&  =  & + 0.003  \pm  0.021 \\
{\cal A}_{\rm CP}^{\rm mix}(B_s\to J/\psi K_{\mathrm S}^0) |_{\rm SM}
& =  & -0.29\phantom{0}   \pm  0.20 .
   \end{array}
\end{equation}
The LHCb collaboration has recently reported the first measurement of these observables 
\cite{LHCb-BspsiKS}:
\begin{equation}
\begin{array}{rcl}
{\cal A}_{\rm CP}^{\rm dir}(B_s\to J/\psi K_{\mathrm S}^0) & = & -0.28 \pm 0.41 {\rm(stat)}  \pm 0.08 
{\rm (syst)} \\
{\cal A}_{\rm CP}^{\rm mix}(B_s\to J/\psi K_{\mathrm S}^0) & = & +0.08 \pm 0.40 {\rm(stat)} 
\pm 0.08 {\rm (syst),} 
\end{array}
\end{equation}
which have still large uncertainties. As discussed in detail in \cite{KR-pen}, the measurement of
CP violation in $B^0_s\to J/\psi K_{\rm S}$ will allow the control of the penguin 
effects in $B^0_d\to J/\psi K_{\rm S}$ at the LHCb upgrade, with additional powerful tests of 
non-factorisable $U$-spin-breaking effects. 
 
 For the Belle II experiment, the $B^0_d\to J/\psi \pi^0$ mode, which is obtained from 
 $B^0_s\to J/\psi K_{\rm S}$  by replacing the strange spectator quark through a down quark, 
 will be an interesting penguin control channel  \cite{CPS}. In this case, exchange penguin 
 annihilation amplitudes have to be neglected in $B^0_d\to J/\psi \pi^0$ as they have no counterpart 
 in $B^0_d\to J/\psi K_{\rm S}$. These effects are expected to be tiny and can be probed through 
 $B^0_s\to J/\psi\pi^0$ and $B^0_s\to J/\psi \rho^0$. There is no evidence for these channels in the 
 current LHCb data.

\subsubsection{The $B^0_s\to J/\psi \phi$, $B^0_d\to J/\psi \rho^0$, $B^0_s\to J/\psi \bar K^{*0}$
System}
In the decay $B^0_s\to J/\psi \phi$, the final state is a mixture of CP-odd and CP-even states, which
have to be disentangled through the angular distribution of the $J/\psi [\to\mu^+\mu^-]\phi [\to\ K^+K^-]$ 
decay products \cite{DDLR,DDF,DFN}. The impact of SM penguin contributions is similar to 
$B^0_d\to J/\psi K_{\rm S}$ \cite{RF-ang,FFM}:
\begin{equation}
A\left(B_s^0\rightarrow (J/\psi \phi)_f\right) = \left(1-\frac{\lambda^2}{2}\right)\mathcal{A}'_f
\left[1+\epsilon a'_fe^{i\theta'_f}e^{i\gamma}\right],
\end{equation}
where the label $f$ distinguishes between the CP-even and CP-odd final state configurations $0,\parallel$ 
and $\perp$, respectively. The penguin parameters $(a'_f,\theta'_f)$ are expected to take different values 
for the different final-state configurations $f$, in contrast to the situation within factorisation \cite{RF-ang}. 
The phases extracted from the CP-violating observables take the following form:
 \begin{equation}
 \phi_{s,f}^{\rm eff} = \phi_s+\Delta\phi_s^{f},
 \end{equation}
 where the hadronic phase shifts $\Delta\phi_s^{f}$ are proportional to 
 the penguin parameters $\epsilon a'_f$ \cite{FFM}.
 
The LHCb collaboration has recently reported the first polarisation-dependent results for 
$\phi_{s,f}^{\rm eff}$ \cite{LHCb-phis-pol}. Within the uncertainties, which are still sizeable, no 
differences between the $\phi_{s,f}^{\rm eff}$ could be revealed. This analysis represents a 
pioneering first step, with interesting future prospects. Assuming 
polarisation-independent values of the penguin parameters yields
\begin{equation}\label{phi_s-eff}
\phi_{s}^{\rm eff}= \phi_s + \Delta\phi_s^{\psi\phi}= -(3.32 \pm  2.81  \pm0.34)^\circ.
\end{equation}

The decay $B^0_d\to J/\psi \rho^0$ is described by the amplitude
\begin{equation}
\sqrt{2} \, A\left(B_d^0\rightarrow (J/\psi \rho^0)_f\right) = - \lambda \mathcal{A}_f
\left[1 - a_fe^{i\theta_f}e^{i\gamma}\right],
\end{equation}
which has a structure that is analogous to those of the $B^0_d\to J/\psi \pi^0$ and 
$B^0_s\to J/\psi K_{\rm S}$ amplitudes. In particular, it shows also 
``magnified" penguin contributions, and allows to control the penguin corrections in $B^0_s\to J/\psi \phi$
\cite{RF-ang}. It should be noted that the hadronic parameters in $B^0_{s,d}\to J/\psi K_{\rm S}^0$ 
and $B^0_d\to J/\psi \rho^0$ are generally expected to differ from one another. 

Following the methods proposed by Zhang and Stone \cite{ZS}, the LHCb collaboration has recently reported
the first measurements for CP violation in the $B^0_d\to J/\psi \rho^0$ channel \cite{LHCb-psi-rho}.
Assuming universal hadronic parameters $a_{\psi\rho}$, $\theta_{\psi\rho}$ and employing
the value of $\phi_d$ in (\ref{a-det}), a $\chi^2$ fit to the data yields the following 
results \cite{KR-pen}:
\begin{equation}
a_{\psi\rho} = 0.037^{+0.097}_{-0.037} \:,\quad \theta_{\psi\rho} 
= -\left(67^{+181}_{-141}\right)^{\circ}\:,
\quad \Delta\phi_d^{J/\psi\rho^0} = -\left(1.5_{-10}^{+12}\right)^{\circ}.
\end{equation}
Using the relations
\begin{equation}
a_{\psi\phi}'
=  \xi a_{\psi\rho},  \quad  \theta_{\psi\phi}'= \theta_{\psi\rho}+\delta \quad 
\end{equation}
with $\xi=1.00\pm0.20$ and $\delta=(0\pm 20)^\circ$ gives the hadronic phase shift
\begin{equation}
\Delta\phi_s^{\psi\phi} = \left[0.08_{-0.72}^{+0.56}\:(\rm{stat})_{-0.13}^{+0.15}\:({\it SU}(3))\right]^{\circ},
\end{equation}
which is consistent with the result obtained in \cite{LHCb-psi-rho}. The comparison with
(\ref{phi_s-eff}) shows that this analysis puts already an impressive constraint on the
hadronic penguin uncertainties. 

Another interesting aspect of the analysis in \cite{KR-pen} is the extraction of the 
hadronic amplitude ratios 
$|\mathcal{A}'_f(B_s\rightarrow J/\psi\phi)/\mathcal{A}_f(B_d\rightarrow J/\psi \rho^0)|$ from
the LHCb data and their comparison with factorisation using form factors from QCD light-cone sum rules 
\cite{BZ}, and a recent PQCD analysis \cite{LWX}. For further details and updates, see \cite{Kristof-PhD}. 
The ratios following from the experimental data do not indicate any large non-factorisable 
$SU(3)$-breaking corrections, thereby supporting the $SU(3)$ assumption when relating 
the $B^0_d\to J/\psi \rho^0$ penguin parameters to their $B^0_s\to J/\psi \phi$ counterparts.

Another penguin probe is offered by the decay $B^0_s\to J/\psi \bar K^{*0}$ \cite{FFM}. In contrast
to $B^0_d\to J/\psi \rho^0$, the final state is flavour-specific and the decay does not exhibit 
mixing-induced CP violation. In \cite{KR-pen}, a new strategy and roadmap to 
control the penguin corrections in the extraction of $\phi_d$ and $\phi_s$ from $B\to J/\psi X$ decays
in the era of Belle II and the LHCb upgrade was presented. In this analysis, there is interesting
interplay between the methods using $B_s$ and $B_d$ decays. Following these lines, the 
expected future impressive experimental errors can eventually be matched by theory.

\subsection{Another Penguin Playground: $B\to D\bar D$ Decays}
This decay system offers yet another laboratory for CP violation and strong interactions. 
In particular, the $B^0_q$--$\bar B^0_q$ mixing phases can be determined through 
CP-violating asymmetries, 
exchange and penguin annihilation topologies can be probed and insights into flavour symmetry
relations can be obtained \cite{RF-psiKS,RF-BDD-07,GRP,JS}. A comprehensive study of the
anatomy of $B\to D\bar D$ decays has recently been performed in \cite{BDD-anatom}. This analysis 
finds enhanced exchange and penguin annihilation topologies, indications for 
significant penguin contributions, discusses factorisation tests employing semileptonic 
$B^0_d\to D_d^-\ell^+\nu_\ell$ modes, and sheds light on the importance of $SU(3)$-breaking effects. 
Moreover, the prospects for the LHCb upgrade and Belle II era were discussed, where a measurement of
the $B^0_s\to D_s^-\ell^+\nu_\ell$ decay would be an interesting new ingredient for the
$B\to D\bar D$ analysis. The detailed implementation of the penguin strategy depends on the values 
of the measured observables, which will eventually allow us to control the penguin effects in the 
determination of $\phi_s$ from the CP asymmetries of the $B^0_s\to D_s^+D_s^+$  channel.

\section{Perspectives for Rare B Decays}
Rare decays of $B$ mesons complement studies of CP violation in a powerful way. In the
SM, these FCNC processes do not receive contributions at the tree level. They offer 
various interesting ways to probe the flavour structure of the SM and to search for footprints of new 
interactions and particles. The field of rare $B$ decays is very rich \cite{BG}. In the following discussion, 
the main focus will be put on the perspectives for rare $B^0_s$ decays and new observables for the LHC
upgrade era.
\subsection{The decay $B^0_s\to\mu^+\mu^-$}
In the SM, this channel receives contributions from penguin and box topologies and is helicity suppressed, 
with a branching ratio proportional to $m_\mu^2$. Consequently, the decay is strongly suppressed
and very rare indeed, which makes the experimental analysis challenging, despite the favourable signature
with two muons in the final state. On the other hand, the hadronic sector is very simple as only
leptons are present in the final state. The whole hadronic sector is described by the $B^0_s$-meson
decay constant $F_{B_{s}}$ parameterising the matrix element 
$\langle 0| \bar b \gamma_5\gamma_\mu s | B^0_s(p)\rangle = i F_{B_s} p_\mu$. The
$B^0_s\to\mu^+\mu^-$ mode belongs therefore to the cleanest and most interesting rare $B$ decays. 

A highlight of LHC run 1 is the observation of the $B^0_s\to\mu^+\mu^-$ decay 
in a combined analysis by the CMS and LHCb collaborations \cite{CMS-LHCb-nature}.
The corresponding branching ratio reads
\begin{equation}\label{CMS-LHCb}
{\mathcal{B}}(B^0_{s}\to\mu^+\mu^-)=(2.8^{+0.7}_{-0.6})\times10^{-9},
\end{equation}
while the most recent theoretical SM analysis gives the following result \cite{Bmumu-TH,bobeth}:
\begin{equation}\label{BRmumu-SM}
{\mathcal{B}}(B^0_{s}\to\mu^+\mu^-)=(3.66\pm0.23)\times10^{-9}.
\end{equation}

Thanks to the sizeable decay width difference $\Delta\Gamma_s$ of the $B^0_s$-meson system, 
a subtle difference arises between the time-integrated ``experimental" branching ratios and their
 ``theoretical" counterparts \cite{BR-paper}. In the latter quantities, the effects from 
 $B^0_s$--$\bar B^0_s$ mixing are ``switched off" by considering decay time $t=0$ \cite{RF-psiKS,RF-ang}.
 In the case of the $B^0_{s}\to\mu^+\mu^-$ decay, this effect at the 10\% level is particularly relevant 
 for the search of NP and has to be properly included when comparing experiment with the 
 SM prediction \cite{Bsmumu-BR}. In fact, (\ref{BRmumu-SM}) takes these effects into 
 account and refers actually to the time-integrated branching ratio, thereby matching the experimental 
 result in (\ref{CMS-LHCb}). 

The combined analysis of the CMS and LHCb data has also resulted in a measurement of the
branching ratio of the $B^0_d\to\mu^+\mu^-$ channel \cite{CMS-LHCb-nature}:
\begin{equation}\label{CMS-LHCb-Bd}
{\mathcal{B}}(B^0_{d}\to\mu^+\mu^-)=(3.9^{+1.6}_{-1.4})\times10^{-10}.
\end{equation}
On the other hand, the SM prediction is given as follows \cite{Bmumu-TH,bobeth}:
\begin{equation}\label{BRmumu-d--SM}
{\mathcal{B}}(B^0_{d}\to\mu^+\mu^-)=(1.06\pm0.09)\times10^{-10}.
\end{equation}
It is useful to consider the ratio ${\cal R}\equiv{\cal B}(B^0_d\to\mu^+\mu^-)/{\cal B}(B^0_s\to\mu^+\mu^-)$ 
\cite{Buras}. In this quantity, the ratio $F_{B_d}/F_{B_s}$  enters, which can even 
be more precisely be calculated by means of lattice QCD than the individual decay constants. The 
SM prediction reads ${\cal R}=0.0295^{+0.0028}_{-0.0025}$ \cite{CMS-LHCb-nature}. Thanks to its 
construction, the quantity ${\cal R}$ takes also in models with MFV the SM value. The fit of the combined
CMS and LHCb data gives ${\cal R}=0.14^{+0.08}_{-0.06}$, which agrees with the SM at the
$2.3\,\sigma$ level \cite{CMS-LHCb-nature}. This interesting experimental situation could indicate
a deviation from the SM and models with MFV. The LHC data could hence be 
first indirect hints of new particles and interactions entering the corresponding Feynman diagrams.
It will be exciting to monitor the future developments.

The measurement of the $B^0_s\to\mu^+\mu^-$ branching ratio at the LHC requires the
use of normalisation channels \cite{FST}:
\begin{equation}
{\mathcal{B}}(B^0_s\to\mu^+\mu^-)
={\mathcal{B}}(B_q\to X)\frac{\epsilon_{X}}{\epsilon_{\mu\mu}}
\frac{N_{\mu\mu}}{N_{X}}\frac{f_q}{f_s}.
\end{equation}
In this relation, where the $N$ are event numbers and the $\epsilon$ factors detector efficiencies,
the ratio of the fragmentations functions $f_s/f_d$ is the major limiting factor. This hadronic 
quantity can be extracted form experimental data. Nevertheless, the $B^0_s\to\mu^+\mu^-$ branching 
ratio cannot be measured with arbitrarily high precision at the LHC upgrade by simply collecting 
more and more statistics. 

The question comes to mind whether there is actually an observable of the rare $B^0_s\to\mu^+\mu^-$
decay beyond the branching ratio that could be exploited at the upgrade era of the LHC. In fact, the
sizeable decay width difference $\Delta\Gamma_s$ provides access to such a new observable 
${\cal A}_{\Delta\Gamma}^{\mu\mu}$ \cite{Bsmumu-BR}, which can be extracted from a 
time-dependent untagged analysis:
\begin{displaymath}
\langle \Gamma(B_s(t)\to \mu^+\mu^-)\rangle\equiv
\Gamma(B^0_s(t)\to  \mu^+\mu^-)+ \Gamma(\bar B^0_s(t)\to  \mu^+\mu^-)
\end{displaymath}
\begin{equation}
\propto e^{-t/\tau_{B_s}}\bigl[\cosh(y_st/ \tau_{B_s})+  {\cal A}_{\Delta\Gamma}^{\mu\mu}
\sinh(y_st/ \tau_{B_s})\bigr],
\end{equation}
where $y_s \equiv \Delta\Gamma_s/2\,\Gamma_s = 0.075 \pm 0.012$, and is encoded in
the effective $B^0_{s}\to\mu^+\mu^-$ lifetime
\begin{equation}
\tau_{\mu^+\mu^-} \equiv \frac{\int_0^\infty t\,\langle \Gamma(B_s(t)\to \mu^+\mu^-)\rangle\, dt}
	{\int_0^\infty \langle \Gamma(B_s(t)\to \mu^+\mu^-)\rangle\, dt}= 
	\frac{\tau_{B_s}}{1-y_s^2}\left[\frac{1+2\,{\cal A}^{\mu\mu}_{\Delta\Gamma}y_s + y_s^2}
	{1 + {\cal A}^{\mu\mu}_{\Delta\Gamma} y_s}\right].
\end{equation}

The general low-energy effective Hamiltonian describing the $B^0_s\to\mu^+\mu^-$ decay
is given by
\begin{equation}
{\cal H}_{\rm eff}=-\frac{G_{\rm F}}{\sqrt{2}\pi} V_{ts}^\ast V_{tb} \alpha
\bigl[C_{10} O_{10} + C_{S} O_S + C_P O_P
+ C_{10}' O_{10}' + C_{S}' O_S' + C_P' O_P' \bigr],
\end{equation}
where in the SM there is only a contribution from the  operator
$O_{10}=(\bar s \gamma_\mu P_L b) (\bar\ell\gamma^\mu \gamma_5\ell)$ with a real 
Wilson coefficient $C_{10}^{\rm SM}$. The observable ${\cal A}_{\Delta\Gamma}^{\mu\mu}$ takes
the following form \cite{Bsmumu-BR}:
\begin{equation}
{\cal A}_{\Delta\Gamma}^{\mu\mu}
=\frac{|P|^2\cos (2\varphi_P-\phi^{\rm NP}_s)-|S|^2\cos(2\varphi_S-\phi^{\rm NP}_s)}{|P|^2+|S|^2}
\,\stackrel{\rm SM}{\longrightarrow}\, 1,
\end{equation}
where
\begin{equation}
P\equiv |P|e^{i\varphi_P}\equiv \frac{C_{10}-C_{10}'}{C_{10}^{\rm SM}}+
{\frac{M_{B_s} ^2}{2\, m_\mu}
\left(\frac{m_b}{m_b+m_s}\right)\left(\frac{C_P-C_P'}{C_{10}^{\rm SM}}\right)}
\,\stackrel{\rm SM}{\longrightarrow}\, 1
\end{equation}
\begin{equation}
S\equiv |S|e^{i\varphi_S}\equiv \sqrt{1-4\frac{m_\mu^2}{M_{B_s}^2}}
{\frac{M_{B_s} ^2}{2\, m_\mu}\left(\frac{m_b}{m_b+m_s}\right)
\left(\frac{C_S-C_S'}{C_{10}^{\rm SM}}\right)}
\,\stackrel{\rm SM}{\longrightarrow}\, 0.
\end{equation}
It is interesting to emphsise that ${\cal A}_{\Delta\Gamma}^{\mu\mu}$ probes also the
CP-violating NP phases $\varphi_P$ and $\varphi_S$.

The observable ${\cal A}_{\Delta\Gamma}^{\mu\mu}$, which is theoretically clean, 
offers a new degree of freedom to probe
NP with the $B^0_s\to\mu^+\mu^-$ decay. It is useful to introduce the following ratio \cite{Bsmumu-BR}:
\begin{equation}
 R  \equiv 
 \frac{{\mathcal{B}}(B^0_{s}\to\mu^+\mu^-)_{\rm exp}}{{\mathcal{B}}(B^0_{s}\to\mu^+\mu^-)_{\rm SM}}
	= \left[\frac{1+{\cal A}^{\mu\mu}_{\Delta\Gamma}\,y_s}{1-y_s^2} \right]  (|P|^2 + |S|^2)
\end{equation}
\vspace*{-0.4truecm}
\begin{displaymath}
= \left[\frac{1+y_s\cos(2\varphi_P-\phi_s^{\rm NP})}{1-y_s^2} \right] |P|^2 + 
\left[\frac{1-y_s\cos(2\varphi_S-\phi_s^{\rm NP})}{1-y_s^2} \right] |S|^2.
\end{displaymath}
The ratio $R$, which takes the current value of $0.82\pm0.21$,  does not allow a 
separation of the $P$ and $S$ contributions. Consequently, NP contributions could still be 
present. Further information from the measurement of 
${\cal A}_{\Delta\Gamma}^{\mu\mu}$ yields 
\begin{equation}
|S|=|P|\sqrt{\frac{\cos(2\varphi_P-\phi^{\rm NP}_s)
-{\cal A}_{\Delta\Gamma}^{\mu\mu}}{\cos(2\varphi_S 
-\phi^{\rm NP}_s)+{\cal A}_{\Delta\Gamma}^{\mu\mu}}},
\end{equation}
thereby allowing us to resolve this situation, as discussed in \cite{Bsmumu-BR}. A detailed analysis
of specific NP scenarios was performed in \cite{BFGK}, showing that interesting regions for NP
in observable space are left by the constraints from the currently available data on rare decays.

The observable ${\cal A}_{\Delta\Gamma}^{\mu\mu}$ -- or equivalently the effective 
$B^0_{s}\to\mu^+\mu^-$ lifetime  -- is a promising new observable for the 
physics agenda of the LHC upgrade. Experimental studies and further theoretical analyses
in NP scenarios are strongly encouraged.

\subsection{New Observables in $B^0_{s}\to\phi\ell^+\ell^-$}
In analogy to the $B^0_s\to \mu^+\mu^-$ decay,  the width difference $\Delta\Gamma_s$ can 
also be exploited in the rare decay $B^0_s\to \phi\ell^+\ell^-$ \cite{DGV}. Here an angular analysis 
is required and the situation is much more involved than in the case of $B^0_s\to \mu^+\mu^-$ 
in view of a much more challenging hadronic situation (form factors, resonances, etc.). 
It is useful to complement the search for NP through $B^0_d\to K^{*0} \mu^+\mu^-$ sketched
below with $B^0_{s}\to\phi\ell^+\ell^-$. Moreover, observables of the time-dependent analysis of the 
angular distribution of the $B^0_d\to K^{*0}(\to \pi^0K_{\rm S})\ell^+\ell^-$ decay offer new analyses
as well. It will be interesting to fully exploit the physics potential of semileptonic rare 
$B_{(s)}$ decays at the Belle II experiment and the LHCb upgrade.

\subsection{Puzzles ins Semileptonic Rare $B$ Decays}
LHCb data for $\bar B^0\to\bar K^{*0}\mu^+\mu^-$ have recently received a lot of 
attention in the theory community \cite{bobeth}, 
in particular the observable $P_5'$ of the angular angular distribution 
\cite{DGMV}--
\cite{DGHMV-pow}. The experimental pattern
may indicate NP effects. However, the theoretical interpretation is challenging in view of hadronic
effects, such as power corrections of the kind $\Lambda_{\rm QCD}/m_b$ affecting form-factor relations
and the impact of hadronic resonances. A detailed analysis has recently been performed in \cite{walt,AS},
finding a sizeable tension between the data and the usual SM interpretation. Interesting future 
prospects are offered by the measurement of the $q^2$ dependence of the $C_9$ Wilson coefficient,
which allows to distinguish between underestimated hadronic effects and NP contributions. 

Another puzzling LHCb measurement has recently also moved into the spotlight (and was 
included in \cite{AS}). It is given by the following ratio \cite{LHCb-RK}:
\begin{equation}
R_K\equiv\frac{{\mathcal{B}}(B\to K\mu^+\mu^-)_{[1,6]}}{{\mathcal{B}}(B\to K e^+e^-)_{[1,6]}}=
0.745^{+0.090}_{-0.074}\pm0.036,
\end{equation}
where the square brackets indicates the range of the $q^2$ bins between (1--6)~$\mbox{GeV}^2$.
This observable offers a test of lepton flavour universality \cite{HK}. In contrast to the
$\bar B^0\to\bar K^{*0}\mu^+\mu^-$ observables, $R_K$ is a very clean quantity, taking the
SM value of one with excellent precision \cite{BHP}. The deviation of the LHCb measurement from the
SM prediction with $2.6\,\sigma$ significance may hence by an indication of the violation of
lepton flavour universality. As the observable is very clean, hadronic effects cannot explain the 
LHCb central value, in contrast to the situation of the $P_5'$ observable. Consequently, it 
could be an experimental fluctuation or a sign of physics beyond the SM which would violate lepton
flavour universality. In the recent literature, 
assuming the exciting latter possibility, various theoretical studies were performed, focussing on 
specific theoretical frameworks such as leptoquarks, $Z'$ models and composite Higgs scenarios. 
For a selection of references, see 
\cite{HMN}--
\cite{AGC}.

In order to get the full picture in the pursuit of NP with the semileptonic rare $B$ decays discussed above, 
the search for lepton-flavour-violating processes is very interesting. Prominent examples of such
decays, which are forbidden in the SM, are $B\to K \mu e$, $B\to K \mu \tau$, $B_s\to \mu e$ and
$B_s\to \mu\tau$. A measurement of any of these modes would be an unambiguous NP signal.

\section{Conclusions and Outlook}
Exciting opportunities for $B$ physics are ahead of us thanks to the recently started run 2 of the LHC,
and to the Belle II experiment and the LHC upgrade in the more distant future. In this presentation, just
a selection out of many interesting topics was covered. There are excellent prospects for measuring 
$\gamma$, with pure decays of the kind $B\to D^{(*)}K^{(*)}$ and $B_s\to D_s^\mp K^\pm$ on the
one hand and $B_s\to K^+K^-$, $B_d\to \pi^+\pi^-$ decays on the other hand, where loop contributions
are involved. New variants of the latter strategy using the $U$-spin symmetry were proposed and LHCb
has presented first pioneering results. Concerning $B\to\pi K$ decays, the focus of $SU(3)$ methods is
changing from the determination of $\gamma$ to probing electroweak penguins, where
$B^0_d\to \pi^0K_{\rm S}$ is a particularly interesting mode for Belle II.

In future high-precision measurements of the $B^0_{d,s}$--$\bar B^0_{d,s}$ mixing phases it will be required
to control penguin corrections. The $B^0_s\to J/\psi K_{\rm S}$ channel, which can be exploited at
the LHCb upgrade, offers the cleanest control of such effects in the determination of $\phi_d$ from 
CP violation in $B^0_d\to J/\psi K_{\rm S}$, while $B^0_d\to J/\psi \pi^0$ will be interesting for Belle II. 
For the determination of $\phi_s$ from $B^0_s\to J/\psi \phi$, the $B^0_d\to J/\psi \rho^0$ mode is the
key player, with first LHCb measurements on CP violation putting already impressive constraints on the
penguin effects and giving valuable insights into $SU(3)$-breaking effects. The $B^0_s\to J/\psi \bar K^{*0}$
channel can also be added to a global analysis of the penguin parameters. A complementary setting
is offered by $B\to D\bar D$ decays.

Rare $B$ decays are currently a particularly exciting topic. The observation 
of $B^0_s\to \mu^+\mu^-$ in a combined analysis of the CMS and LHCb data is a highlight 
of run 1 of the LHC. It will be interesting to monitor also the situation for $B^0_d\to \mu^+\mu^-$, 
where the current LHC data indicate a tension with respect to the SM and NP models with MFV. 
The effective lifetime of the $B^0_s\to \mu^+\mu^-$
offers a new theoretically clean observable to complement the branching ratio measurement, which 
should be added to the physics agenda of the LHC upgrade. The LHCb data for the
$\bar B^0\to\bar K^{*0}\mu^+\mu^-$ channel and the measurement of the 
$R_K={\mathcal{B}}(B^+\to K^+\mu^+\mu^-)/{\mathcal{B}}(B^+\to K^+e^+e^-)$  observable have
led to a lot of interest in the community. While hadronic effects may be underestimated in the
former case, the latter observable is very clean and would indicate a violation of lepton flavour
universality. Further intriguing patterns arise in the data for $B\to \tau \nu$ and $B\to D^{(*)}\tau \nu$ 
decays. Concerning the latter mode, new results were presented at this conference by the 
LHCb \cite{LHCb-SLtau} and Belle \cite{Belle-SLtau} collaborations. These ``anomalies" have led 
to a large number of theoretical studies and speculations in the context with physics beyond the SM.

In the future, it will hopefully be possible to resolve the discrepancy between the determinations
of $|V_{ub}|$ and $|V_{cb}|$ from inclusive and exclusive semileptonic $B$ decays (see \cite{PDG} for
an overview). This long-standing problem affects the determination of the side $R_b$ of the 
UT, which is -- together with the angle $\gamma$ -- a key parameter for the SM prediction of 
$\sin2\beta$ (see \cite{KR-pen} for a recent illustration). The LHCb collaboration has recently 
made the first measurement of $|V_{ub}|$ from a baryonic decay, which is given by 
$\Lambda_b^0\to p\mu^-\bar\nu_\mu$ \cite{LHCb-Vub-Lam}. This determination exploits progress in 
lattice QCD calculations of the corresponding from factors, and is in agreement with the average 
value of $|V_{ub}|$ extracted from exclusive semileptonic $B$ decays. 

It will be crucial for the full exploitation of $B$ physics in the next decade to have continued strong
interactions between theory and experiment. Key topics are related to the impact of strong interactions
and hadronic physics, with issues related to the factorisation of hadronic matrix elements of
non-leptonic decays, $SU(3)$-breaking corrections in decay amplitude relations, and the optimal 
use of data to shed light on these issues. It is also important to think further about new observables 
to test the flavour sector of the SM  and to further explore correlations and patterns between various  
processes in specific NP models. The latter studies will strongly benefit from the future results for 
direct NP searches by the ATLAS and CMS experiments. 

Looking at the current $B$-decay data, it seems that $(2\mbox{--}3)\sigma$ deviations from the 
SM are accumulating. The crucial question is whether we are eventually revealing signs of new 
particles and interactions, or whether these effects will disappear with more sophisticated experimental
and theoretical analysis. Interesting times with exciting prospects for $B$ physics are ahead of us!

\vspace*{0.5truecm}

\noindent
{\it Acknowledgements}\\
I would like to thank Toru Iijima and his colleagues for the excellent organisation of FPCP 2015 which 
made my visit to Nagoya most enjoyable. I would also like to thank my colleagues and students for 
the pleasant collaboration on various topics presented in this talk.
\end{document}